\documentclass{scrartcl}                     
\usepackage{graphicx}
\usepackage[utf8]{inputenc}

\usepackage{amsmath,amsfonts,bm,amssymb,amscd} 
\DeclareMathAlphabet{\mathcalON}{OT1}{pzc}{m}{n}
\newcommand{\s}{{\mathrm s}}
\newcommand{\dd}{{\mathrm d}}
\newcommand{\ddt}{\frac{\dd}{\dd t}}
\DeclareBoldMathCommand{\cA}{{\cal A}}
\DeclareBoldMathCommand{\cB}{{\cal B}}
\newcommand{\cP}{\mathbf{p}}
\DeclareBoldMathCommand{\cQ}{{\cal Q}}
\DeclareBoldMathCommand{\cC}{{\cal C}}
\newcommand{\cW }{\mathbf{w}}
\DeclareBoldMathCommand{\cI}{{\cal J}}
\newcommand{\Tlim}{\mathcalON{T}}
\newcommand{\bfa}{\mathbf{a}}
\newcommand{\bfe}{\mathbf{e}}
\newcommand{\bfw}{\mathbf{w}}
\newcommand{\bfh}{\mathbf{h}}
\newcommand{\bfr}{\mathbf{r}}
\newcommand{\bfP}{\mathbf{P}}
\newcommand{\bfc}{\mathbf{c}}
\newcommand{\bfx}{\mathbf{x}}
\newcommand{\bfA}{\mathbf{A}}
\newcommand{\bfM}{\mathbf{M}}
\newcommand{\bfK}{\mathbf{K}}
\newcommand{\bfB}{\mathbf{B}}
\newcommand{\bfJs}{\mathbf{J}_{\mathrm{s}}}
\newcommand{\bfv}{\mathbf{v}}
\newcommand{\cBtil}{\bm{\widetilde{\cB}}}
\newcommand{\cCtil}{\bm{\widetilde{\cC}}}
\newcommand{\bfxhat}{\mathbf{\widehat{x}}}
\newcommand{\bfchat}{\mathbf{\widehat{c}}}

\newcommand{\xhat}{{\widehat{x}}}
\newcommand{\Np}{N_{\mathrm p}}
\newcommand{\Nppwmbal}{N_{\mathrm{p},\mathrm{pwmbal}}}
\newcommand{\Ns}{N_{\mathrm s}}
\newcommand{\Ts}{T_{\mathrm s}}
\newcommand{\fs}{f_\mathrm{s}}

\newcommand{\iL}{i_\mathrm{L}}
\newcommand{\vC}{v_{\mathrm C}}
\newcommand{\vi}{v_{\mathrm i}}
\newcommand{\RL}{R_{\mathrm{L}}}

\newcommand{\Ltwo}{{\mathrm{L}^2}}

\usepackage{pdfcomment}
\usepackage[normalem]{ulem}
\usepackage{authblk}

\title{Multirate PWM balance method for the efficient field-circuit coupled simulation of power converters}
\author[1,2,3]{Andreas Pels}
\author[1,2]{ Herbert De Gersem}
\author[3]{Ruth V. Sabariego}
\author[1,2]{Sebastian Schöps}
{
\affil[1]{\small Graduate School of Computational Engineering, Technische Universität Darmstadt, Dolivostraße 15, 64293 Darmstadt, Germany}
\affil[2]{Institut für Teilchenbeschleunigung und Elektromagnetische Felder, Technische Universit\"at Darmstadt, Schlossgartenstraße 8, 64289 Darmstadt, Germany}
\affil[3]{Department of Electrical Engineering, EnergyVille, KU Leuven, Kasteelpark Arenberg 10, 3001 Leuven, Belgium}
}
\date{}
\begin{document}
\maketitle

\begin{abstract}
  The field-circuit coupled simulation of switch-mode power converters with conventional time discretization is computationally expensive since very small time steps are needed to appropriately account for steep transients occurring inside the converter, not only for the degrees of freedom (DOFs) in the circuit, but also for the large number of DOFs in the field model part. An efficient simulation technique for converters with idealized switches is obtained using multirate partial differential equations, which allow for a natural separation into components of different time scales. This paper introduces a set of new PWM eigenfunctions which decouple the systems of equations and thus yield an efficient simulation of the field-circuit coupled problem. The resulting method is called the multirate PWM balance method.
\end{abstract}

\section{Introduction}
\begin{figure}[t]
  \centering
  \includegraphics[width=0.8\columnwidth]{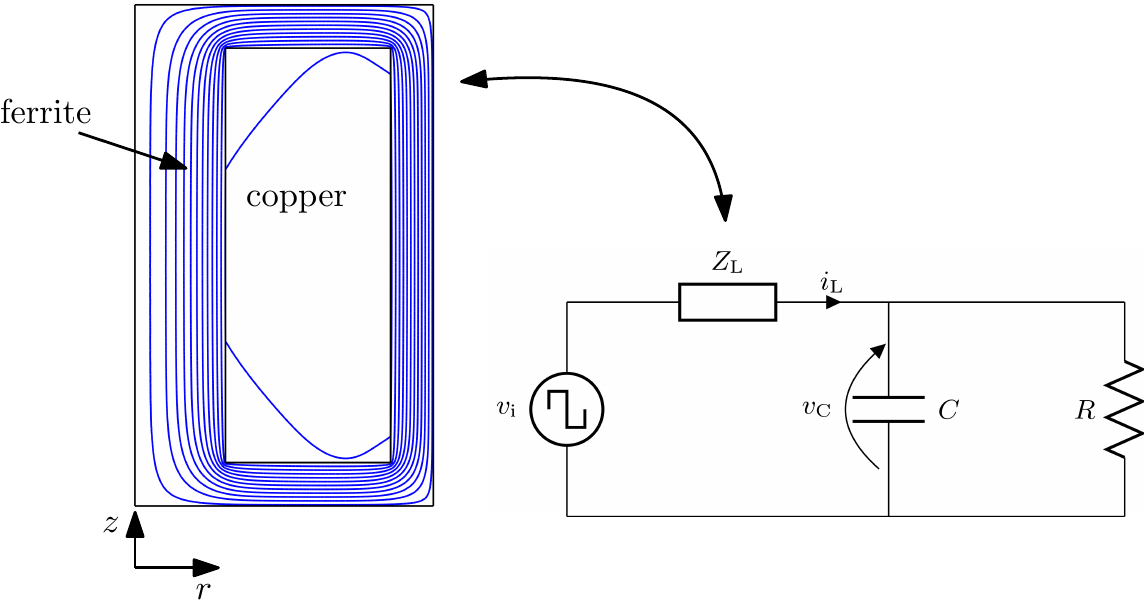}
  \caption{Simplified circuit of the buck converter in continuous conduction mode with $C=10\,\text{$\mu$F}$ and $R=30\,\Omega$. The field model of the pot inductor (axisymmetric around $z$-axis) is designed to have an inductance of $L=65\,\text{mH}$ and a series resistance of $\RL=800\,\text{m$\Omega$}$ at DC. The figure shows the equipotential lines of the magnetic vector potential. } 
  \label{fig:buckCircuit}
  \end{figure}

Switch-mode power converters are used in various devices from small-scale applications like mobile phone chargers to industrial large-scale applications like welding devices \cite{Mohan_2003aa}. These converters use transistors to switch on and off the input voltage to produce an output voltage, which, in average, has the desired amplitude. A filter circuit is used to smoothen the output. The simulation of these devices is computationally expensive since, through the transistor switching, steep transients occur in the converter. Furthermore often a switch-event detection is necessary to avoid step size rejection or even solver failures \cite{Tant_2018aa}. A multirate method has been developed in \cite{Pels_2018aa,Pels_2019aa} which uses the concept of Multirate Partial Differential Equations (MPDEs) \cite{Brachtendorf_1996aa,Roychowdhury_2001aa} and a combination of a Galerkin ansatz and conventional time discretization to efficiently solve problems with pulsed excitation. The method is applicable to power converters in which the switching behavior is idealized and known a-priori. It is particularly efficient in the case of linear elements. 
Some circuit elements may only be accurately represented by field models. For example the induced currents in the conducting materials of an inductor usually cause eddy current losses, which can easily be accounted for in a field model but not in a circuit model. In this paper the multirate method from \cite{Pels_2018aa,Pels_2019aa} is applied to a linear buck converter circuit (see Fig.~\ref{fig:buckCircuit}) in which the inductor is represented by a 2D finite element model. This substantially increases the size of the strongly coupled system of equations. To still ensure an efficient simulation, a basis transformation is applied to the pulse-width modulation (PWM) basis functions \cite{Gyselinck_2013ab} leading to decoupled systems of equations which can be solved efficiently in parallel. The resulting method is called the multirate PWM balance method in analogy with the harmonic balance method where harmonic functions take the place of the PWM basis functions. Numerical results on the buck converter show the efficiency and accuracy of the proposed method in field-circuit coupled problems. 

The paper is structured as follows. Section \ref{sec:multirate} introduces the concept of MPDEs and explains the solving procedure using Galerkin approach and conventional time discretization. Subsequently section \ref{sec:PWMBFs} presents the original PWM basis functions as described in \cite{Gyselinck_2013ab}. In section \ref{sec:trafoPWMBF} the PWM eigenfunctions are developed and their advantageous properties for the solving process are highlighted. Finally section \ref{sec:numRes} summarizes numerical results and compares the three different solution approaches, i.e., conventional time discretization and the MPDE approach with PWM basis functions on the one hand and PWM eigenfunctions on the other hand. The paper is concluded by summarizing its content in section \ref{sec:conclusion}.

\section{Multirate formulation}\label{sec:multirate}
Let the field-circuit coupled model \cite{Schops_2013aa} of the converter be described by the system of ordinary differential or differential-algebraic equations
\begin{equation}
  \begin{aligned}
    \bfA \ddt \bfx(t)+\bfB \bfx(t) &= \bfc(t), 
  \end{aligned}
  \label{equ:DAEOriginal}
\end{equation}
where $\bfA \in \mathbb{R}^{\Ns \times \Ns}$ is a possibly singular matrix, $\bfB\in\mathbb{R}^{\Ns \times \Ns}$ is assumed to be a regular matrix, $\bfx(t)\in\mathbb{R}^{\Ns}$ is the unknown solution, $\bfc(t)\in\mathbb{R}^{\Ns}$ is the excitation, and $t\in(0, \Tlim]$ is the simulation interval. The initial state of the model is given by consistent \cite{Lamour_2013aa} initial values $\bfx(0)=\bfx_0$. The ideal pulsed excitation
\begin{equation}
  \vi(t)=\left\{
  \begin{array}{ll}
    V_0 & \text{for all } \tau(t) \leq D \\
    0 & \text{otherwise}
  \end{array}\right.,
  \label{equ:pulsedVoltage}
\end{equation}
is used as input of the power converter circuit. We denote by $\tau(t) = \frac{t}{\Ts} \text{ modulo } 1$ the relative time, $\Ts$ is the switching cycle and $D$ is the duty cycle.

The system of Multirate Partial Differential(-Algebraic) Equations (MPDEs or MPDAEs) with two time scales corresponding to \eqref{equ:DAEOriginal} is given by \cite{Brachtendorf_1996aa,Roychowdhury_2001aa,Pulch_2007ac}
\begin{equation}
  \bfA \, \left(\frac{\partial \bfxhat}{\partial t_1} + \frac{\partial \bfxhat}{\partial t_2}\right) + \bfB \,\bfxhat(t_1,t_2) = \bfchat(t_1,t_2) \, ,
  \label{equ:MPDEOriginal}
\end{equation}
where $\bfxhat(t_1, t_2)$ is the unknown multivariate solution and $\bfchat(t_1,t_2)$ is the multivariate excitation. Choosing the multivariate excitation such that $\bfchat(t,t)=\bfc(t)$, the solution of \eqref{equ:DAEOriginal} and \eqref{equ:MPDEOriginal} are related by $\bfxhat(t,t)=\bfx(t)$. Thus, the solution of \eqref{equ:DAEOriginal} can be calculated solving the MPDEs and extracting the solution along a diagonal through the computation domain. To solve the MPDEs, additional conditions need to be specified. For the present application, a combination of initial and boundary values is applied. Initial values are supplied by $\bfxhat(0,t_2)=\bfh(t_2)$, i.e., at $t_1=0$, where $\bfh$ with $\bfh(0)=\bfx_0$ is a function defining the initial values for all $t_2$. The solution along the fast time scale $t_2$ is periodic, i.e., $\bfxhat(t_1, t_2+\Ts) = \bfxhat(t_1, t_2)$. The multivariate right-hand side is chosen as $\bfchat(t_1, t_2)=\bfc(t_2)$, i.e, the pulses of the excitation occur along the fast time scale. It is possible to use MPDEs with more than two time scales. However, in the applications of this paper, it is not necessary and furthermore often not feasible since the dimension of the computation domain increases and thus also the computational effort to calculate the solution. 

To solve the MPDEs \eqref{equ:MPDEOriginal}, a Galerkin approach and time discretization is applied \cite{Pels_2019aa,Bittner_2014aa}. The $j$-th solution component $\xhat_j(t_1, t_2)$ is approximated by expanding it into periodic basis functions $p_k$ depending on the fast time scale $t_2$ and coefficients $w_{j,k}$ depending on the slow time scale $t_1$ 
\begin{equation}
  \xhat_j^h(t_1, t_2) := \sum_{k=0}^{\Np} w_{j,k}(t_1) \, p_k(\tau(t_2)) \,,
  \label{equ:solExp}
\end{equation}
where the periodicity of the basis functions is accounted for by using the relative time $\tau(t_2) \ = \ \frac{t_2}{T_\s} \ \text{modulo} \ 1.$
Applying the Galerkin approach with respect to $t_2$ and over one period of the excitation $[0,\Ts]$ leads to
\begin{equation}
  \cA\, \frac{\dd  \cW}{\dd t_1} + \cB \, \cW(t_1) \ = \ \cC(t_1) \, ,
  \label{equ:ReducedMPDESystem}
\end{equation}
with block matrices $\cA=\cI\otimes\bfA, \quad \cB=\cI\otimes\bfB+\cQ\otimes\bfA$, where
\begin{gather}
  \cI  =  \Ts \int\limits_0^1  \mathbf{\bar{p}}(\tau) \, \cP^\top \!(\tau)  \,\dd \tau, \,\, \cQ  =  - 
  \int\limits_0^1  \frac{\dd \mathbf{\bar{p}}(\tau)}{\dd \tau } \, \cP^\top\!(\tau)  \,  \dd \tau  \, ,
  \label{equ:matricesIQ}
\end{gather}
and right-hand side
\begin{equation}
  \cC(t_1)=\int\limits_{0}^{T_\s} \mathbf{\bar{p}}(\tau(t_2)) \otimes \bfchat(t_1,t_2)   \dd t_2 \,.
  \label{equ:matricesC}
\end{equation}
$\mathbf{\bar{p}}$ denotes the complex conjugate of $\cP$ and $\otimes$ denotes the Kronecker product.

\section{PWM basis functions}\label{sec:PWMBF}
\label{sec:PWMBFs}
\begin{figure}[t]
  \centering
  \includegraphics{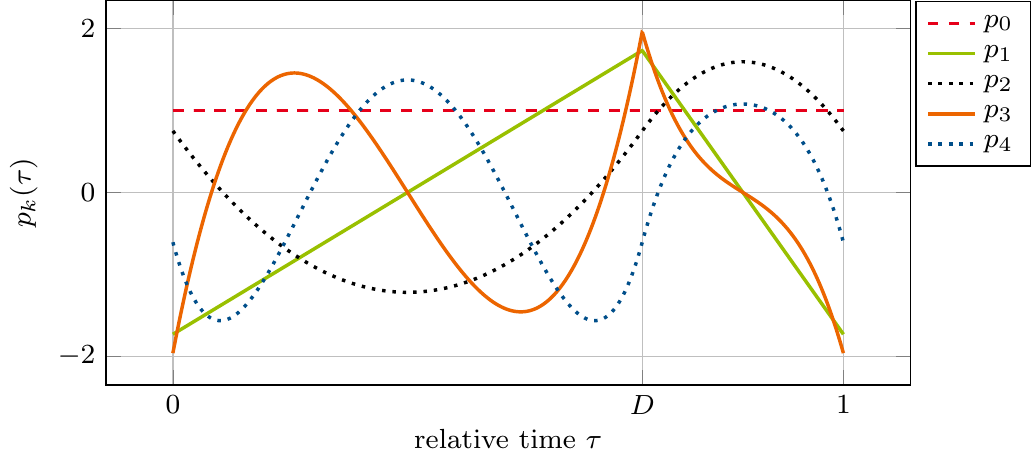}
  \caption{Original PWM basis functions $p_k(\tau)$, $k\in \{0,1,2,3,4\}$.}
  \label{fig:PWMBFs}
\end{figure}

The PWM basis functions developed in \cite{Gyselinck_2013ab} are built up starting from the zero-th constant basis function $p_0(\tau)=1$ and the piecewise linear basis function 
\begin{equation}
  \renewcommand*{\arraystretch}{1.3}
  \displaystyle
  p_1(\tau) \ = \ \left\{  
  \begin{array}{ll} 
    \sqrt{3} \ \frac{2\tau-D}{D} \quad & \mathrm{if} \ \ \ 0 \leq \tau \leq D \\ 
    \sqrt{3} \ \frac{1+D-2\tau}{1-D}  \quad & \mathrm{if} \  \ \ D \leq \tau \leq 1    \end{array}
  \right. ,
  \label{equ:PWMbasis_p1}
\end{equation}
which includes the duty cycle $D$ of the excitation by construction. The higher-order basis functions $p_k(\tau)$, $2\leq k \leq \Np$ are recursively obtained by integrating the basis functions of lower order $p_{k-1}(\tau)$ and orthonormalizing them using the Gram-Schmidt algorithm. The generated basis functions are depicted in Fig.~\ref{fig:PWMBFs}. 

For the PWM basis functions, the matrices $\cI$ and $\cQ$ from \eqref{equ:matricesIQ} are given by $\Ts$ multiplied by the identity matrix (due to the orthonormality of the basis functions) and a square matrix with around 25\,\% of non-zero entries, respectively. Solving the problem requires time stepping of the entire system \eqref{equ:ReducedMPDESystem}.

\section{PWM eigenfunctions}\label{sec:trafoPWMBF}
\label{sec:complexPWMBFs}
\begin{figure}
  \centering
  \includegraphics{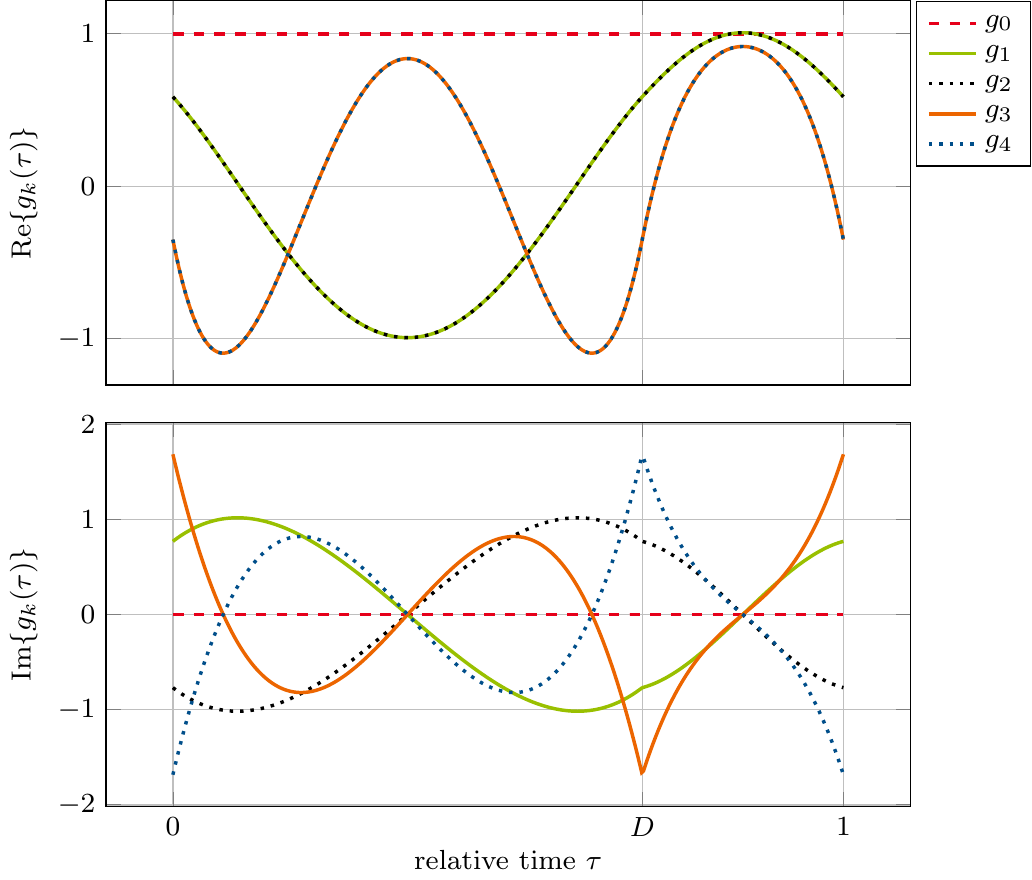}
  \caption{PWM eigenfunctions $g_k(\tau)$, $k\in \{0,1,2,3,4\}$, i.e., $\Np=4$. (top) real part. (bottom) imaginary part.}
  \label{fig:newPWMBFs}
\end{figure}

To enable an easy parallelization of the method, the equations \eqref{equ:ReducedMPDESystem} can be decoupled, for example by diagonalizing $\cQ$, i.e., a basis transformation. We define new basis functions $g_k(\tau)$ as linear combinations of the PWM basis functions, i.e., 
\begin{equation}
  g_k(\tau):=\sum\limits_{l=0}^{\Np} v_{k,l} \, p_l(\tau),
\end{equation}
where $v_{k,l}$ are unknown coefficients with $k\in\{0, \dots, \Np\}$, and $g_k(\tau)$ are eigenfunctions of the time derivative operator
\begin{equation}
  \frac{\dd}{\dd \tau} \,g_k(\tau) = \lambda_k \, g_k(\tau).
\end{equation}
We enforce this property in a weak sense by a Galerkin approach, i.e.,
\begin{equation}
  -\int\limits_{0}^{1} g_k(\tau) \frac{\dd p_m(\tau)}{\dd \tau} \dd \tau \; = \lambda_k \int\limits_{0}^{1} g_k(\tau) p_m(\tau) \dd \tau,
  \label {equ:galTrafoBFs}
\end{equation}
where integration by parts and the periodicity of the basis functions is used. 
Inserting the expansion of the basis functions into \eqref{equ:galTrafoBFs} gives
\begin{equation}
  \Ts \cQ \bfv_k = \lambda_k \cI \bfv_k.
\end{equation}
Since $\cI$ is $\Ts$ multiplied by the identity matrix (thanks to the orthonormality of the PWM basis functions), the $\lambda_k$ and $\bfv_k$ are the eigenvalues and eigenvectors of the matrix $\cQ$, respectively. Furthermore since $\cQ$ is real-valued and  skew symmetric, and therefore a normal matrix,
the eigenvectors $\bfv_k$ are orthonormal. The new basis functions (complex-valued) are depicted in Fig.~\ref{fig:newPWMBFs} for $\Np=4$. Note that the basis consists of pairs of conjugate complex basis functions. 

Inserting the transformed basis functions instead of the PWM basis functions into \eqref{equ:matricesIQ} and \eqref{equ:matricesC} leads, using the orthonormality of the eigenvectors, to 
\begin{equation}
  \cA\, \frac{\dd  \cW}{\dd t_1} + \cBtil \, \cW(t_1) \ = \ \cCtil(t_1) \, ,
  \label{equ:ReducedMPDESystemTrafo}
\end{equation}
where $\cA$ as in \eqref{equ:ReducedMPDESystem},\begin{align}
  \cBtil&=\cI\otimes\bfB+\boldsymbol{\Lambda}\otimes\bfA,\label{equ:matricesABTrafo}\\
  \cCtil(t_1)&=\int\limits_{0}^{T_\s} \mathbf{\bar{g}}(\tau(t_2)) \otimes \bfchat(t_1,t_2)   \dd t_2 \, ,  \label{equ:matricesCTrafo}
\end{align}
and $\boldsymbol{\Lambda}$ is a diagonal matrix with diagonal entries ${\lambda_0, \lambda_1, \dots, \lambda_{\Np}}$. Thus the resulting matrices in \eqref{equ:ReducedMPDESystemTrafo} are block-diagonal and the degrees of freedom can be block-wisely decoupled. This leads to $\Np+1$ independent systems of equations given by
\begin{equation}
	\Ts \bfA \frac{\dd \bfw_{k}}{\dd t_1} + (\Ts \bfB + \lambda_k \bfA) \bfw_{k} = \int\limits_{0}^{T_\s} \bar{g}_k(\tau(t_2)) \,  \bfchat(t_1,t_2)   \, \dd t_2\, \text{ for } k = 0,\dots,\Np,
	\label{equ:decoupledEquations}
\end{equation}
where $\bfw=[\bfw_{0}^\top, \bfw_{1}^\top, \dots, \bfw_{\Np}^\top]^\top$
Note that if a diagonal entry in $\boldsymbol{\Lambda}$ is complex, there is also a complex conjugate counterpart. The solutions of the decoupled system of equations corresponding to this complex eigenvalue and its conjugate complex counterpart, are, as a result, complex conjugate to each other. Therefore it is sufficient to solve one of them. This is similar to harmonic balance methods in which the harmonic basis functions are given by pairs of complex conjugates leading to similar systems of equations. In analogy to ``harmonic balance method'', we call the developed method the ``multirate PWM balance method''.  

\section{Test case and numerical results}\label{sec:numRes}
The method is applied to the buck converter from Fig.~\ref{fig:buckCircuit}, where the pot inductor is represented by a 2D field model with conducting core material (ferrite, $\sigma_{\mathrm{fe}}=250\,$S/m). The coils are modeled as stranded conductors. The simulation interval is given by $\Psi=[0,10]\,$ms. The switching frequency is $\fs=\frac{1}{\Ts}=1000\,$Hz. For the pulsed excitation \eqref{equ:pulsedVoltage} we use $V_0=24\,$V. All calculations are performed in MATLAB. The partial differential equations governing the magnetoquasistatic problem are given by
\begin{equation}
  \sigma(\bfr) \frac{\partial \bfA_{\mathrm{m}}(\bfr, t)}{\partial t} + \nabla \times \left((\mu(\bfr)^{-1}) \nabla \times \bfA_{\mathrm{m}}(\bfr, t)\right) = \bfJs(\bfr, t),
\end{equation}
where $\bfr$ is the position vector, $t$ is the time, $\bfA_{\mathrm{m}}$ is the modified magnetic vector potential \cite{Emson_1988aa}, $\bfJs$ are the imposed currents, $\mu=4 \pi \times 10^{-7}\,$H/m is the magnetic permeability and $\sigma$ is the conductivity which is only non-zero in the ferrite core ($\sigma_{\mathrm{fe}}$). The problem is considered on a 2D planar domain with homogeneous Dirichlet boundary conditions.

Correspondingly, the Finite Element magnetoquasistatic \cite{Schops_2013aa} discretization of the magnetoquasistatic inductor model is given by the differential-algebraic system of equations \cite{Nicolet_1996aa}
\begin{equation}
  \bfM_{\sigma} \ddt \bfa(t) +\bfK \bfa(t) = \bfP \iL(t),
\end{equation}
where $\bfM_{\sigma}$ is the singular conductivity matrix, $\bfK$ is the stiffness matrix, $\bfa(t)$ gathers the degrees of freedom (DOFs) related to the magnetic vector potential, $\bfP$ is the discretization of the winding function \cite{Schops_2013aa} and $\iL(t)$ is the current through the inductor. 
The field-circuit coupling is expressed as follows. An additional variable is introduced for the magnetic flux linkage
$\Phi(t)=\bfP^\top \bfa(t)$. All equations are coupled monolithically into the index-1 differential-algebraic system of equations \cite{Bartel_2011aa}
\begin{align}
  \bfP^\top \bfa - \Phi &= 0, \\
  C \ddt \vC - \iL + \frac{1}{R} \vC &= 0, \\
  \bfM_{\sigma} \ddt \bfa - \bfP \iL + \bfK \bfa &= \mathbf{0}, \\
  \ddt\Phi + \RL \iL + \vC &= \vi(t),
\end{align}
which for the example in Fig.~\ref{fig:buckCircuit} contains a total of $11053$ DOFs. The initial conditions are given by $\vC(0)=0$, $\iL(0)=0$ and $\bfa(0)=\mathbf{0}$.

The initial condition for the MPDEs \eqref{equ:MPDEOriginal} can be written as 
\begin{equation}
	h_j(t_2) \approx \sum_{k=0}^{\Np} w_{j,k}(0) \, p_k(\tau(t_2))
\end{equation}
where $h_j$ is the $j$-th element of $\bfh$. It only has to satisfy the condition $\bfh(0) = \bfxhat(0,0)=\bfx_0$. Consequently there is a high degree of freedom in choosing the initial values $\bfw(0)$ for the system of equations \eqref{equ:ReducedMPDESystem}. However not all choices lead to an efficient simulation, i.e., low dynamics in the slow time scale. 
The following choice of initial values has proven advantageous. 
First, the steady-state solution is calculated, i.e., 
\begin{equation}
	\bfw^s= \cB^{-1}\cC(0). 
\end{equation}
Secondly, the initial coefficients for $k=1,\dots, \Np$ are extracted from the steady-state solution $w_{j,k}(0)=w^s_{j,k}$ for all $j$. The remaining coefficients are calculated by solving the solution expansion \eqref{equ:solExp} for $w_{j,0}(0)$ and using the condition $\bfxhat(0,0)=\bfx_{0}$. In summary the initial coefficients are given by
\begin{equation}
  w_{j,k}(0)=\left\{
	\begin{array}{ll}
		w^s_{j,k} & \text{for }k=1,\dots,\Np \text{ and for all } j=1,\dots,\Ns \\
		x_j(0) - \sum\limits_{l=1}^{\Np} w^s_{j,l} p_l(0)	& \text{for }k=0 \text{ and for all } j=1,\dots,\Ns.
	\end{array}\right.
  \label{equ:initialValues}
\end{equation}
The initial conditions for the system of equations \eqref{equ:ReducedMPDESystemTrafo} are computed similarly using the PWM eigenfunctions. 
Other choices of initial values may still lead to the correct solution but might impair the efficiency of the method.

To calculate the reference solution with a conventional adaptive time discretization, the MATLAB solver \texttt{ode15s} is used. It is modified to restart the simulation at the known switching instances. Consistent initial values for the restart of the solver are calculated by using a Newton-Raphson algorithm to solve the set of algebraic equations. The required differential variables are taken from the solution at the end of the prior solution interval. After finding the new set of initial values, the initial slopes of the differential variables are calculated by solving the subsystem of ordinary differential equations for the slope. 

\begin{figure}
  \centering
  \includegraphics{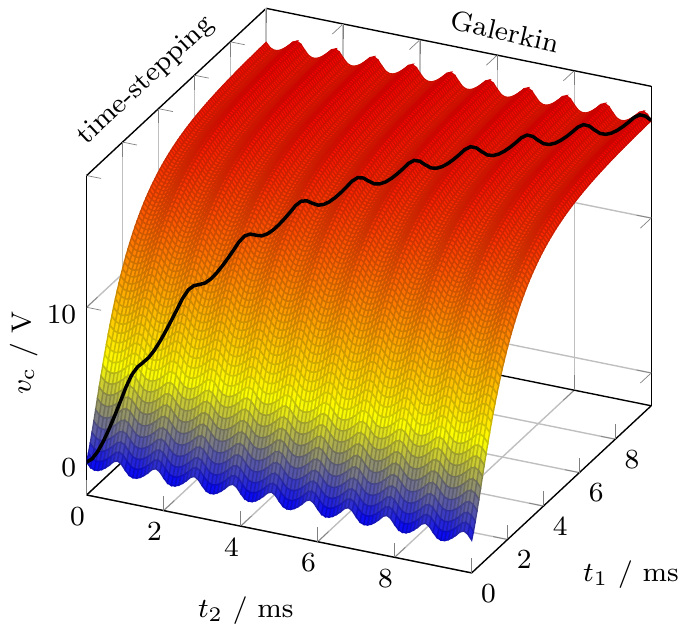}
  \caption{Multivariate voltage at the capacitor calculated using the multirate PWM balance method. The solution component corresponding to the original system of equations \eqref{equ:DAEOriginal} is extracted along a diagonal and marked as a black curve.}
  \label{fig:MPDE3D}
\end{figure}

\begin{figure}[t]
  \centering
  \includegraphics{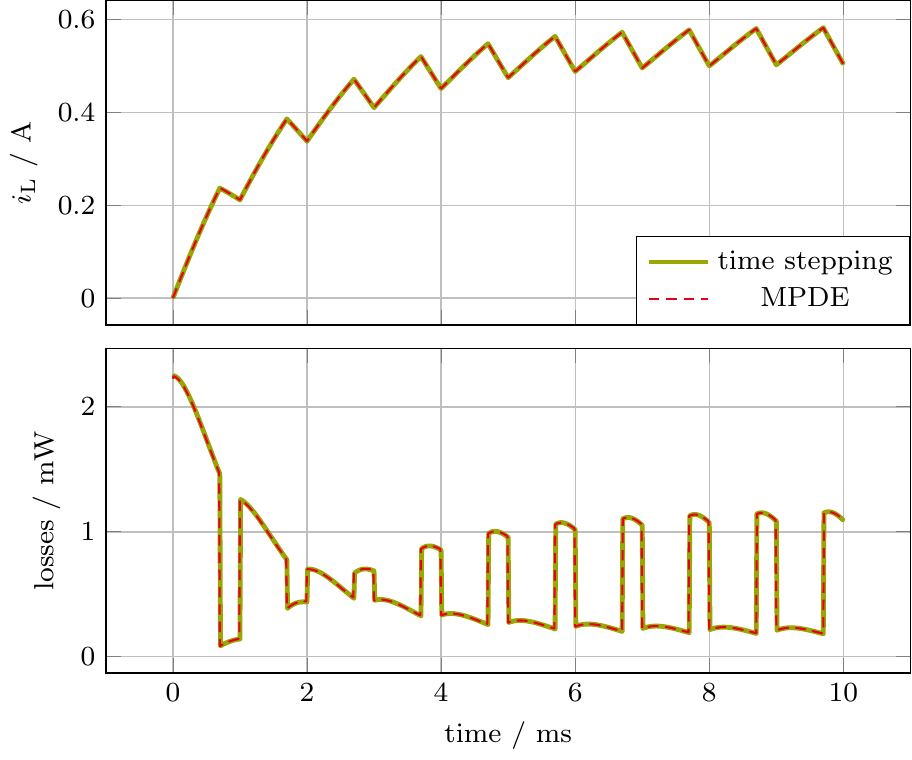}
  \caption{(top) Reference solution calculated using conventional adaptive time discretization compared to the solution obtained by the MPDE approach with $\Nppwmbal=4$ PWM eigenfunctions. The relative $\Ltwo$ error of the current through the inductor similar to \eqref{equ:l2erroreps} is approximately $3 \times 10^{-5}$. (bottom) Joule losses in the core material due to eddy currents.}
  \label{fig:compMPDEAna}
\end{figure}

\begin{figure}[t]
  \centering
  \includegraphics{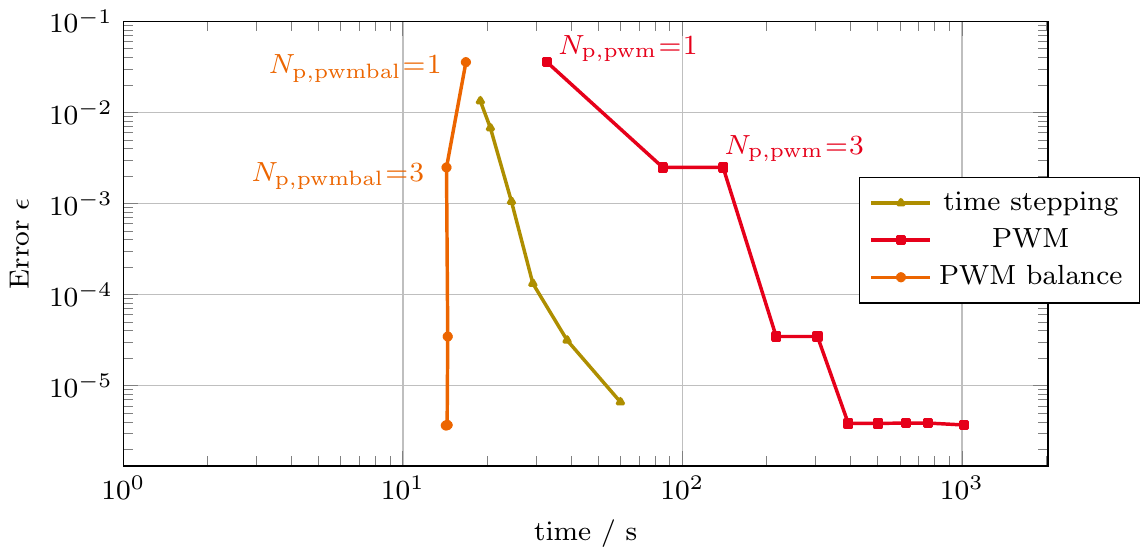}
  \caption{Error $\epsilon$ as defined in \eqref{equ:l2erroreps} over time for solving the systems of equations. The MPDE approach with PWM eigenfunctions (multirate PWM balance method) is considerably faster than the MPDE approach with the original PWM basis functions and the conventional time discretization.}
  \label{fig:errorOverTime}
  \end{figure}

\begin{figure}[t]
  \centering
  \includegraphics{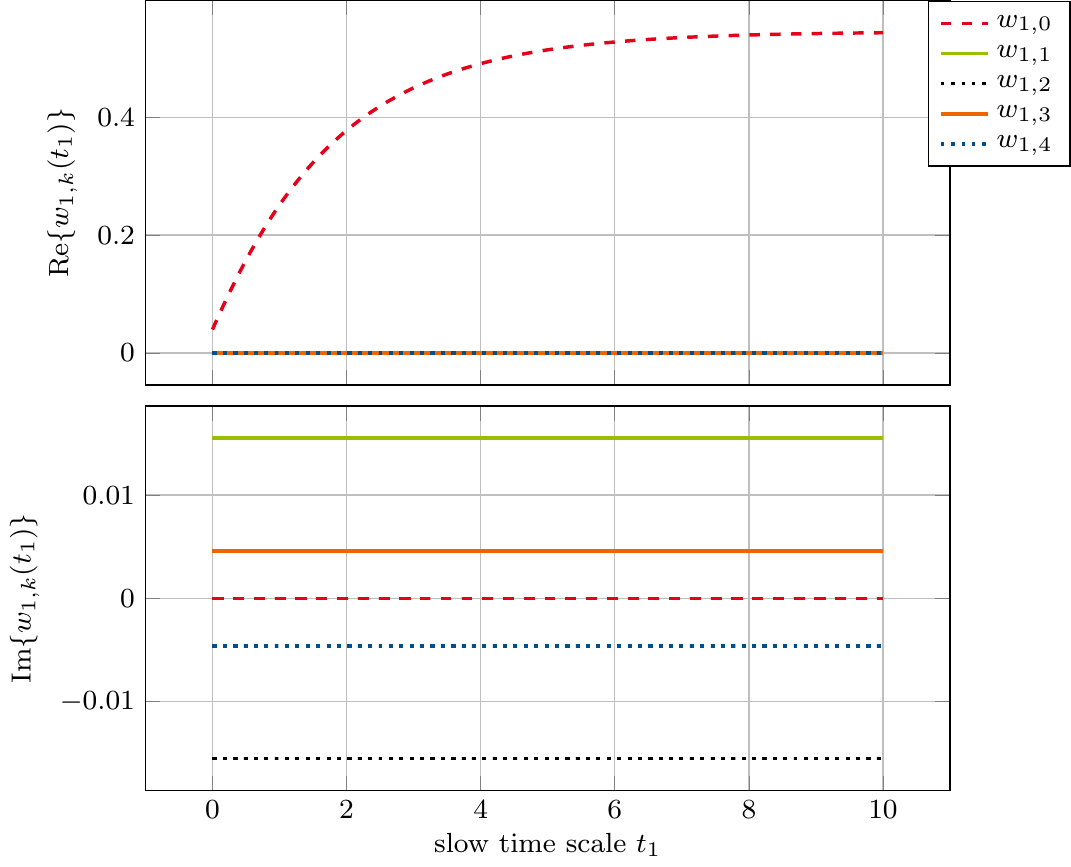}
  \caption{Coefficients $w_{1,k}$ for the inductor current calculated by solving \eqref{equ:ReducedMPDESystemTrafo} with $\Nppwmbal=4$. (top) real part. The coefficients $w_{1,1}, \dots, w_{1,4}$ are approximately the same therefore they are hard to distinguish visually. (bottom) imaginary part.}
  \label{fig:coefficients}
  \end{figure}

The multivariate solution $\bfxhat(t_1,t_2)$ calculated using the multirate PWM balance method, i.e., solving \eqref{equ:ReducedMPDESystemTrafo} with \texttt{ode15s}, is reconstruced using \eqref{equ:solExp} and the multivariate voltage at the capacitor is depicted in Fig.~\ref{fig:MPDE3D}. The corresponding solution component of the original system of equations \eqref{equ:DAEOriginal} is extracted along a diagonal through the computation domain. 
Fig.~\ref{fig:compMPDEAna} shows the current through the inductor along with the reference solution. The agreement between the multirate PWM balance method solution and the reference solution is excellent. The Joule losses in the core material due to eddy currents are calculated by
\begin{equation}
  P_\mathrm{eddy}(t)= \int\limits_{\Omega}  \mathbf{E}(\bfr, t) \cdot \sigma(\bfr) \mathbf{E}(\bfr, t) \, \dd \bfr = \left(\bfe(t)^\mathrm{H}\right) \, \bfM_{\sigma}  \, \bfe(t),
\end{equation}
where $\mathbf{E}$ is the electric field strength, $\Omega$ is the spatial computation domain, the superscript $\mathrm{H}$ denotes the Hermitian, i.e., the complex conjugate transposed, and $\bfe(t)=-\ddt \bfa(t)$ is the line-integrated discrete electric field. The Joule losses are plotted as well in Fig.~\ref{fig:compMPDEAna}. Fig.~\ref{fig:coefficients} depicts the solution of \eqref{equ:ReducedMPDESystemTrafo}, i.e., the coefficients $\bfw(t_1)$, exemplary for the current through the inductor $\iL$. As one can see, using the initial values \eqref{equ:initialValues}, only the coefficient $w_{j,0}$ corresponding to the zero-th basis function varies and the others stay constant. To quantify the accuracy and efficiency of the multirate PWM balance method, it is compared to conventional time discretization and to the MPDE approach with the original PWM basis functions. Different settings are considered: To analyze the performance of the conventional time discretization, the relative and absolute tolerance setting of the solver is changed, i.e., $\mathrm{abstol}=\mathrm{reltol} \in [10^{-6}, 10^{-1}]$; For the case of the multirate PWM balance method and the MPDE approach with the original PWM basis functions, relative and absolute tolerances are fixed at $\mathrm{abstol}=\mathrm{reltol}=10^{-7}$ and the number of basis functions $\Np \in [1,10]$ is changed. The accuracy is measured for the voltage output of the converter, i.e., the voltage at the capacitor. The relative $\Ltwo$ error is given by
\begin{equation}
  \epsilon(\mathrm{tol}, n) = \frac{||v_\mathrm{C,ref}(t)-v_\mathrm{C}^h(\mathrm{tol},n,t)||_{\Ltwo(\Psi)}}{||v_\mathrm{C,ref}(t)||_{\Ltwo(\Psi)}},
  \label{equ:l2erroreps}
\end{equation}
where $v_\mathrm{C,ref}$ is the reference solution and $v_\mathrm{C}^h$ is the solution using the multirate PWM balance method and the MPDE approach with the original PWM basis functions. The norm is approximated using mid-point quadrature. 
Fig.~\ref{fig:errorOverTime} shows the error plotted as a function of the solution time, i.e., the time that \texttt{ode15s} needs. For conventional time discretization the time for solving consists of the time that is needed to calculate consistent initial values and slopes after switching, and the actual time that \texttt{ode15s} needs. The time to calculate consistent initial values and slopes depends on the number of switching instants and is thus constant if the switching frequency or simulation interval does not change. It is given by approximately 16 seconds. The total time displayed in Fig.~\ref{fig:errorOverTime} is the sum of both contributions. 

As one can see the MPDE approach with the original PWM basis functions is considerably slower than conventional time stepping. This is due to the fact that the already large systems of equations \eqref{equ:DAEOriginal} (due to field-circuit coupling) are even further increased in size through the Galerkin approach. The stagnation of the error at $10^{-6}$ in Fig.~\ref{fig:errorOverTime} for values larger than $\Np=7$ is caused by the chosen accuracy of \texttt{ode15s}. Furthermore one can see that when adding another basis function the error does decrease with every second basis function. This was already observed in \cite{Pels_2019aa,Gyselinck_2013ab}. For this reason the error for the PWM eigenfunctions is only plotted for $\Nppwmbal=1,2,4,6,8,10$. Since the systems of equations resulting from the multirate PWM balance method are decoupled, they can be solved efficiently in parallel. For each basis function $g_k$ with $k=0,\dots,\Np$, a complex-valued initial value problem of the form \eqref{equ:decoupledEquations} has to be solved. The size of these systems of equations is the same as that of the original system of equations \eqref{equ:DAEOriginal}. 
However, the time for solving is considerably smaller since less time steps are necessary for the same solution accuracy. Note that due to the choice of the initial values \eqref{equ:initialValues} most coefficients in \eqref{equ:ReducedMPDESystemTrafo} for this numerical example do not change and only those corresponding to the zero-th basis function vary. This means that only the decoupled system of equations which corresponds to the zero-th basis function takes considerable computational effort to solve. In a parallel computing environment one would choose as many processor cores as basis functions ($\Np+1$). The overall runtime is then determined only by the initial value problem that takes the longest to integrate. For this numerical example it is $k=0$. The communication overhead between processors is not taken into account since it is highly implementation and machine dependent. The slightly decreasing time to solution when $\Nppwmbal>1$ is owed to the fact that initial values according to \eqref{equ:initialValues} take more a-priori information into account which leads to smaller number of time steps and faster simulation. The overall accuracy of the method is problem-specific and always depends on both the tolerance for the solver and the number of basis functions. An a-priori determination of the number of basis functions and the solver tolerance is not yet available. An a-posteriori estimator can be constructed by increasing the number of basis functions and comparing the solutions. The resulting error is also related to the time stepping error.

The MPDE approach works also for nonlinear problems. However, similarly as for the harmonic balance case, the decoupling is not straightforward anymore. Furthermore the PWM basis functions and thus also the PWM eigenfunctions might not be able to represent the solution of problems with nonlinear elements \cite{Pels_2019aa}. If the amplitude of the ripples is small compared to the amplitude of the envelope, the particular efficient approach described in \cite{Pels_2018aa} can be applied. It uses only the slowly varying envelope to evaluate the nonlinearities. Although the assembly of the field model matrices for a new envelope can not be parallelized, the matrices in \eqref{equ:ReducedMPDESystemTrafo} can still be decoupled and calculations to obtain the following time step can be run in parallel.

\section{Conclusion}\label{sec:conclusion}
A new efficient technique was presented for field-circuit coupled models of DC-DC power converters, in which the switches are idealized and the filtering circuit is linear. The already existing MPDE technique with PWM basis functions splits the solution into fast varying and slowly varying parts. In this paper this method has been improved by introducing a new set of PMW basis functions which decouple the systems of equations similar as in the harmonic balance method. The new method, now called multirate PWM balance method, enables a parallel solution of all PWM modes resulting in a speed-up amounting to a factor 4 for the test example. 

\section{Acknowledgements}
This work is supported by the ``Excellence Initiative'' of German Federal and State Governments and the Graduate School CE at TU Darmstadt.
The authors thank Johan Gyselinck for fruitful discussions. Further thanks go to Jonas Bundschuh and Erik Sk\"ar for their contribution to the first implementation of the PWM eigenfunctions.

\bibliography{abbrv,english,library}
\bibliographystyle{ieeetr}      
\end{document}